\documentclass[a4paper,12pt]{article}
\usepackage{psfrag}
\usepackage{amssymb}
\usepackage{a4wide,cite,epsfig}

\parskip\smallskipamount
\newcommand{\tpol}{\mathchoice{{2\pi\over L}}{2\pi/L}{2\pi/L}{2\pi/L}}

\newcommand{\oh}{{1\over2}}

\begin{document}
\begin{titlepage}
\begin{flushright}
SHEP-0516\\
\end{flushright}
\bigskip
\begin{center}
{\bf
{\Large
A Numerical Study of Partially Twisted Boundary Conditions}\\[2.5ex]
{\large J.M.~Flynn, A.~J\"uttner and C.T.~Sachrajda}\\
(UKQCD Collaboration)}\\[2ex]
{\sl
School of Physics and Astronomy, University of Southampton,\\
Southampton, SO17 1BJ, UK.}

\bigskip
\bigskip{\large\bf Abstract}\\[3ex]
\parbox[t]{0.85\textwidth}{
We investigate the use of \textit{partially twisted} boundary
conditions in a lattice simulation with two degenerate flavours of
improved Wilson sea quarks. The use of twisted boundary conditions on
a cubic volume ($L^3$) gives access to components of hadronic momenta
other than integer multiples of $\tpol$. Partial twisting avoids the
need for new gluon configurations for every choice of momentum, while,
as recently demonstrated, keeping the finite-volume errors
exponentially small for the physical quantities investigated in this
letter. In this study we focus on the spectrum of pseudo scalar and
vector mesons, on their leptonic decay constants and on $Z_P$, the
matrix element of the pseudo scalar density between the pseudo scalar
meson and the vacuum. The results confirm the momentum shift imposed
by these boundary conditions and in addition demonstrate that they do
not introduce any appreciable noise. We therefore advocate the use of
partially twisted boundary conditions in applications where good
momentum resolution is necessary.}
\end{center}
\bigskip
\hspace{1.2cm}PACS: 
11.15 Ha, % Lattice gauge theory
12.38 Gc % Lattice QCD calculations\\
\end{titlepage}

%%%%%%%%%%%%%%%%%%%%%%
\section{Introduction}
%%%%%%%%%%%%%%%%%%%%%%
\vspace{-.2cm}

In lattice simulations of QCD on a cubic volume ($V=L^3$) with
periodic boundary conditions on the fields, the components of
hadronic momenta $p_i$ are quantized in integer multiples of $\tpol$.
For currently available lattices this implies that the lowest non-zero
momentum is large, typically $500\,\mathrm{MeV}$ or so, and there are
large gaps between neighbouring momenta. This limits the
phenomenological reach of simulations, particularly for momentum
dependent quantities such as the form-factors of weak semileptonic
decays of hadrons. In ref.~\cite{Bedaque:2004kc} Bedaque proposed the
use of \textit{twisted} boundary conditions\footnote{See the
references cited in~\cite{Bedaque:2004kc} and~\cite{deDivitiis:2004kq}
for earlier related ideas.} for the quark fields $\psi$
\begin{equation}
\label{eq:twistdef}
\psi(x_i+L)=e^{i\theta_i}\,\psi(x_i)\,.
\end{equation}
Twisted boundary conditions allow for simulations with arbitrary
components of hadronic momenta. For example, the momentum of a meson
composed of a quark with flavour $1$ satisfying boundary conditions
with a twisting angle $\vec{\theta}_1 =
(\theta_{11},\theta_{12},\theta_{13})$ and an antiquark of flavour $2$
with angle $\vec{\theta}_2$\, is
\begin{equation}\label{eq:momspectrum}
\vec{p}=\tpol\,\vec{n}-\frac{\vec{\theta}_1- \vec{\theta}_2}{L}\,,
\end{equation}
where $\vec{n}$ is a vector of integers.

The practical difficulty in using twisted boundary conditions in
lattice simulations with dynamical quarks is that it requires the
generation of a new set of gauge field configurations for every choice
of twisting angle(s). In refs.~\cite{Sachrajda:2004mi,Bedaque:2004ax}
it was shown that for many physical quantities one can use
\textit{partially twisted} boundary conditions, i.e. impose twisted
boundary conditions for the valence quarks but periodic boundary
conditions for the sea quarks, thus eliminating the need for new
simulations for every choice of momentum and making the technique
practicable. The physical quantities for which partially twisted
boundary conditions can be applied include those with at most a single
hadron in the initial and final states (and possibly even in
intermediate states), for which the finite-volume effects decrease
exponentially with the volume. For these processes the finite-volume
effects depend on the twisting angle(s) but remain exponentially
small.

For some processes with energies above a two-body threshold, such
as $K\to\pi\pi$ decays with the two-pions in an isospin zero
state, the finite-volume effects decrease only as powers of the
volume and must be subtracted for acceptable precision to be
reached. We are not able to perform these subtractions if
partially twisted boundary conditions are used. Here we will only
consider processes for which such a problem does not arise.

In this letter we confirm the theoretical results of
ref.~\cite{Sachrajda:2004mi} in a numerical study of partially twisted
boundary conditions for dynamical, non-perturbatively improved Wilson
fermions. In particular we find that:
\begin{itemize}
\item The energies of $\pi$ and $\rho$-mesons (with masses below the
  two-pion threshold) satisfy the expected dispersion relation
  \begin{equation}
  \label{eq:disprel}
  E_{\pi,\,\rho}^2 =
    m_{\pi,\,\rho}^2+\left(\vec{p}_{\rm lat}-{\vec\theta_1 -
    \vec\theta_2 \over L}\right)^2,
  \end{equation}
  where $\vec\theta_1$ and $\vec\theta_2$ are the twisting angles of
  the two valence quarks and $\vec{p}_\mathrm{lat}=(\tpol)\vec{n}$ is
  the contribution to the meson's momentum introduced by the Fourier
  transform of the correlation function.

  This study extends the one in ref.~\cite{deDivitiis:2004kq} where
  the dispersion relation for pseudo scalar mesons with twisted
  boundary conditions in the quenched approximation was found to be
  consistent with expectations.
\item The values of the leptonic decay constants of $\pi$ and $\rho$
  mesons and of the matrix element $\langle 0|P|\pi\rangle$ of the
  pseudo scalar density $P$ are independent of the twisting angles as
  expected.
\end{itemize}

A further reassuring result of our study is that twisted boundary
conditions do not introduce additional noise in the data. As we
increase the meson's momentum by suitably varying the angles
$\vec\theta_{1,2}$, the statistical errors on meson masses and matrix
elements increase smoothly. However, when comparing results obtained
with twisted and periodic boundary conditions with similar momenta
(i.e. momenta close to $\tpol$ or $\sqrt{2}(\tpol)$) the errors are
found to be comparable.

The plan of the remainder of this letter is as follows. In the next
section we present the details of our computation, the parameters of
the simulation (including the choice of twisting angles) and a
description of the analysis. We present our results in
sec.~\ref{sec:results} and conclusions in sec.~\ref{sec:concs}.

\vspace{-.5cm}
%%%%%%%%%%%%%%%%%%%%%%%%%%%%%%%%%%%%%%%%%%%%%%%%
\section{Details of the Simulation and Analysis}
%%%%%%%%%%%%%%%%%%%%%%%%%%%%%%%%%%%%%%%%%%%%%%%%
\vspace{-.2cm}

We study meson observables on sets of gauge configurations which were
generated with two degenerate flavours of sea quarks using
non-per\-tur\-ba\-tively improved Wilson fermions and the plaquette
gauge action on the torus with periodic boundary conditions
($\beta=5.2$, $a\approx0.1\,\mathrm{fm}$, $c_{\rm SW}=2.0171$,
$(L/a)^3\times T/a =16^3\times 32$). We used the ensembles of field
configurations which were studied in detail
in~\cite{Allton:2001sk,Allton:2004qq} and took over the suggested
separation of measurements by $40$ trajectories in our analysis. The
simulated quark masses are summarized in table~\ref{tab_sim-params}. 
Propagators and correlators were calculated using the FermiQCD
libraries~\cite{DiPierro:2000bd,DiPierro:2001yu,DiPierro:2003sz}. 
We stress that the aim of the present study is to investigate the
consistency and effectiveness of using partially twisted boundary
conditions at fixed values of the quark mass. We do not attempt to
perform a chiral extrapolation.
\begin{table}
\begin{center}
\begin{tabular}{ccc}
\hline\hline\\[-1.5ex]
$\kappa_\mathrm{val}=\kappa_\mathrm{sea}$ & $m_\pi/m_\rho$ &
 $N_\mathrm{meas}$\\[0.5ex]
\hline\\[-2ex]
0.13500 & 0.697(11) & 200\\
0.13550 & 0.566(16) & 200\\[0.2ex]
\hline \hline
\end{tabular}\caption{Simulation parameters.}\label{tab_sim-params}
\end{center}
\vspace{-.8cm}
\end{table}

For each flavour of valence quark we impose the boundary
conditions in eq.~(\ref{eq:twistdef}) for a variety of twisting
angles $\vec{\theta}=(\theta_1,\theta_2,\theta_3)$. When
evaluating the corresponding propagators we make use of the change
of quark field variables
\begin{equation}
\psi(x) = e^{i{\vec{\theta}\cdot\vec{x} \over
L}}\tilde{\psi}(x),
\end{equation}
where $\tilde{\psi}(x)$ satisfies periodic boundary conditions. The
phase factor cancels in all terms of the lattice fermion action except
for the spatial hopping terms which now become (for $i=1,2,3$)
\begin{equation}
\overline{\tilde{\psi}}(x)
 \left[ e^{i{a\theta_i\over L}} U_i(x)(1-\gamma_i)\tilde{\psi}(x+\hat i)
      + e^{-i{a\theta_i\over L}}U^\dagger_i(x-\hat i)
           (1+\gamma_i)\tilde{\psi}(x-\hat i)
 \right].
\end{equation}
In practice therefore, the partially twisted quark propagator can be
computed by inverting the standard improved Wilson-Dirac operator in a
gauge field background where the link variables \{$U_i(x)$\} have been
replaced by \{$e^{i{a\theta_i\over L}}U_i(x)$\}.

The physical observables which we study in this letter are the
energies and leptonic decay constants of the pseudo scalar and vector
mesons and the matrix element of the pseudo scalar density. In order to
determine these, we compute the following correlation
functions:
\begin{eqnarray}
C_{A_0 P}(t,\vec{p}) &=& \sum\limits_{\vec x}
  e^{i\vec p_\mathrm{lat}\cdot\,\vec x}
  \langle 0|A^I_0(\vec x,t)P^\dagger(0)|0\rangle\,,\label{corrA0P}\\
C_{PP}(t,\vec{p}) &=& \sum\limits_{\vec x}
  e^{i\vec p_\mathrm{lat}\cdot\,\vec x}
  \langle 0|P(\vec x,t)P^\dagger(0)|0\rangle\,,\label{corrPP}\\
C_{A_0 A_0}(t,\vec{p}) &=& \sum\limits_{\vec x}
  e^{i\vec p_\mathrm{lat}\cdot\,\vec x}\langle 0|A_0^I(\vec x,t)
  (A_0^I(0))^\dagger|0\rangle\,,\label{corrA0A0}\\
C_{V_i V_i}(t,\vec{p}) &=& \sum\limits_{\vec x}
  e^{i\vec p_\mathrm{lat}\cdot\,\vec x}\langle 0|V_i^I(\vec
  x,t)(V_i^I(0))^\dagger|0\rangle \quad\mbox{(no sum on $i$)}
  \label{corrViVi}\,,
\end{eqnarray}
where $P(x)$ is the pseudo scalar density
\begin{equation}\label{eq:pdef}
P(x)=\overline\psi_2(x)\gamma_5\psi_1(x)
\end{equation}
for quarks of flavour $1$ and $2$ (with twisting angles
$\vec{\theta}_1$ and $\vec{\theta}_2$), and $V^I_\mu(x)$ and
$A^I_\mu(x)$ are the improved vector and axial-vector currents
\begin{eqnarray*}
V^I_\mu(x)&=&\overline\psi_2(x)\gamma_\mu\psi_1(x)+ac_V(g_0)\,
   \oh(\partial^\ast_\nu+\partial_\nu)
   \overline\psi_2(x)\sigma_{\mu\nu}\psi_1(x)\label{eq:videf}\\
A^I_\mu(x)&=&\overline\psi_2(x)\gamma_\mu\gamma_5\psi_1(x)+a
c_A(g_0)\oh(\partial^\ast_\mu+\partial_\mu)P(x)\,.\label{eq:aidef}
\end{eqnarray*}
Here, $\partial_\mu$ and $\partial^\ast_\mu$ are the forward and
backward derivatives and $c_V(g_0)$ and $c_A(g_0)$ are improvement
coefficients which we take from~\cite{Harada:2002jh}
and~\cite{DellaMorte:2005se} respectively. Since we are primarily
interested in the effects of twisted boundary conditions we do not
attempt to compute the renormalization constants of $P$, $V_\mu$
and $A_\mu$, nor do we implement improvement factors of the form
$1+b(g_0)m_qa$, where $m_q$ is the mass of the quark.
The inclusion of these factors would of course be necessary if we were
attempting to determine the physical leptonic decay constants.
However, they are overall factors for each choice of quark mass and
are independent of the twisting angles, while it is precisely the
dependence on these angles which is the object of our study.

The momentum, $\vec{p}$, of the meson is given by
\begin{equation}
\vec p = \vec p_\mathrm{lat} - {\vec \theta_1 - \vec
\theta_2\over L}\,,
\end{equation}
where $p_\mathrm{lat}=(\tpol)\vec{n}$ and $\vec{n}$ is a vector of
integers.

At large values of $t$ the time dependences of
(\ref{corrA0P})--(\ref{corrViVi}) approach:
\begin{eqnarray}
C_{A_0 P}(t,\vec{p}\,)\hspace{-2mm}&\to&\hspace{-2mm}{1\over
E_{\pi}}\,Z_P\,M_0(\vec{p}\,)\,
e^{- E_{\pi}\, T/2} \sinh((t-T/2)E_{\pi}),\label{CAP}\\
C_{PP}(t,\vec{p}\,)\hspace{-2mm}&\to&\hspace{-2mm}{1\over
E_{\pi}}\,Z_P^2\,
e^{-E_{\pi}\,T/2} \cosh((t-T/2)E_{\pi})\label{CPP},\\
C_{A_0 A_0}(t,\vec{p}\,)\hspace{-2mm}&\to&\hspace{-2mm}{1\over
E_{\pi}}\,M_0^2(\vec{p}\,)\,
 e^{-E_{\pi}\,T/2} \cosh((t-T/2)E_{\pi})\label{CAA},\\
C_{V_i V_i}(t,\vec{p}\,)\hspace{-2mm}&\to&\hspace{-2mm}{1\over
E_{\rho}}\,N^2_{i}(\vec p\,)\,e^{-E_{\rho}\,T/2}
\cosh((t-T/2)E_{\rho})\label{CVV}\;\quad(i=1,2,3),
\end{eqnarray}
where, for each choice of quark masses, we have denoted the lightest
pseudo scalar and vector mesons by $\pi$ and $\rho$ respectively and
$E_{\pi}$ and $E_{\rho}$ are the corresponding energies which we
expect to satisfy the dispersion relations in eq.~(\ref{eq:disprel}).
The notation for the matrix elements is as follows:
\begin{eqnarray}
Z_P &=& \langle 0|P(0)|\pi(\vec{p}\,)\rangle, \label{Z_P}\\
M_0(\vec{p}\,) &=&
  \langle 0\,|\,A_0(0)\,|\,\pi(\vec{p}\,)\rangle = f_\pi E_\pi,\label{M_0}\\
N^2_{i}(\vec{p}\,)&=& \sum_{\lambda} |\langle
  \,0|\,V_i(0)\,|\,\rho(\vec{p},\lambda)\rangle|^2 = f_\rho^2
  m_\rho^2 \left(1+{p_{i}^2\over m_\rho^2}\right)\label{N_ii}\,
\end{eqnarray}
where the index $\lambda$ labels the $\rho$-meson's polarization
state.

In this letter we study the validity of the dispersion relation in
eq.~(\ref{eq:disprel}) and the independence of $f_\pi$, $f_\rho$ and
$Z_P$ of the momentum. We evaluate the quark propagators for four
values of the twisting angle $\vec\theta$:
\begin{equation}\label{eq:thetas}
\vec{\theta} = \vec{0},\  (2,0,0),\ (0,\pi,0)\ \textrm{and}\
(3,3,3)\,.
\end{equation}
For each value of $\kappa_\mathrm{val}$, quark and antiquark
propagators with all possible pairs $\vec{\theta}_1$ and
$\vec{\theta}_2$ were combined to construct correlation functions
for mesons with a variety of momenta. Moreover we also combined
them with Fourier momenta $\vec{p}_\mathrm{lat} =
(0,\pm\,\tpol,0)$ to increase the range of momenta which can be
reached. When presenting our results in the following section, we
include for comparison results without twisting
($\vec\theta_1=\vec\theta_2=0$), obtained by averaging over the
$12$ equivalent momenta with
$|\vec{p}_{\mathrm{lat}}|=\sqrt{2}\times \tpol$ and those obtained
by averaging over the eight equivalent momenta with
$|\vec{p}_{\mathrm{lat}}|=\sqrt{3}\times\tpol$. Of course this
averaging reduces the statistical errors and this should be borne
in mind when comparing the errors at these untwisted momenta with
those at momenta with $\vec\theta_1-\vec\theta_2\neq\vec0$ for
which such averaging is not possible.

Applying the jackknife procedure to the data for the correlation
functions in eqs.~(\ref{corrA0P})--(\ref{corrA0A0}), we have extracted
all observables in the pseudo scalar channel from a combined
non-linear $\chi^2$ fit to the functional form suggested by
(\ref{CAP})--(\ref{CAA}), (\ref{Z_P}) and (\ref{M_0}). The fit-ranges
were chosen to yield compatible results under variation of the range
by at least one unit in $t/a$. 
We applied the same procedure in the vector channel, combining the
data for the correlation function~(\ref{corrViVi}) for $i=1,2$ and $3$
in one fit using the expressions in eqs.~(\ref{CVV}) and
(\ref{N_ii}).

\vspace{-.5cm}
%%%%%%%%%%%%%%%%%%%%%%%%%%%%%%%%%%%%%%%%
\section{Results}\label{sec:results}
%%%%%%%%%%%%%%%%%%%%%%%%%%%%%%%%%%%%%%%%
\vspace{-.2cm}

\begin{figure}
\begin{center}
 \begin{minipage}{.495\linewidth}
   \psfrag{kappa}[c][t][1][0]{$m_\pi/m_\rho=0.70$}
   \psfrag{pLsq}[t][c][1][0]{$(\vec{p}L)^2$}
   \psfrag{aEthetasq}[b][b][1][0]{$(aE_{\pi/\rho})^2$}
  \epsfig{scale=.275,angle=-90,file=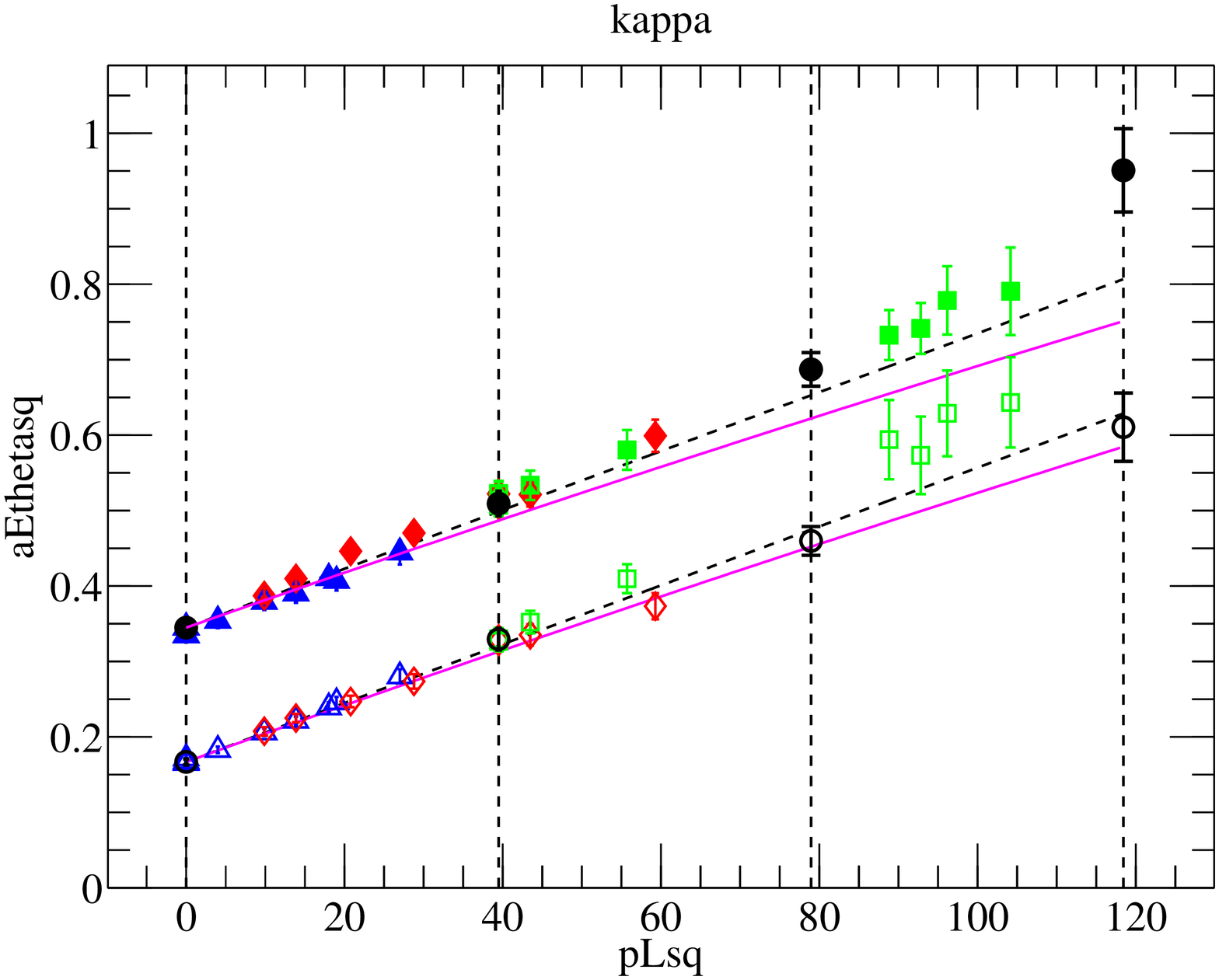}
 \end{minipage}
 \begin{minipage}{.495\linewidth}
   \psfrag{kappa}[c][t][1][0]{$m_\pi/m_\rho=0.57$}
   \psfrag{pLsq}[t][c][1][0]{$(\vec{p}L)^2$}
   \psfrag{aEthetasq}[b][b][1][0]{$(aE_{\pi/\rho})^2$}
  \epsfig{scale=.275,angle=-90,file=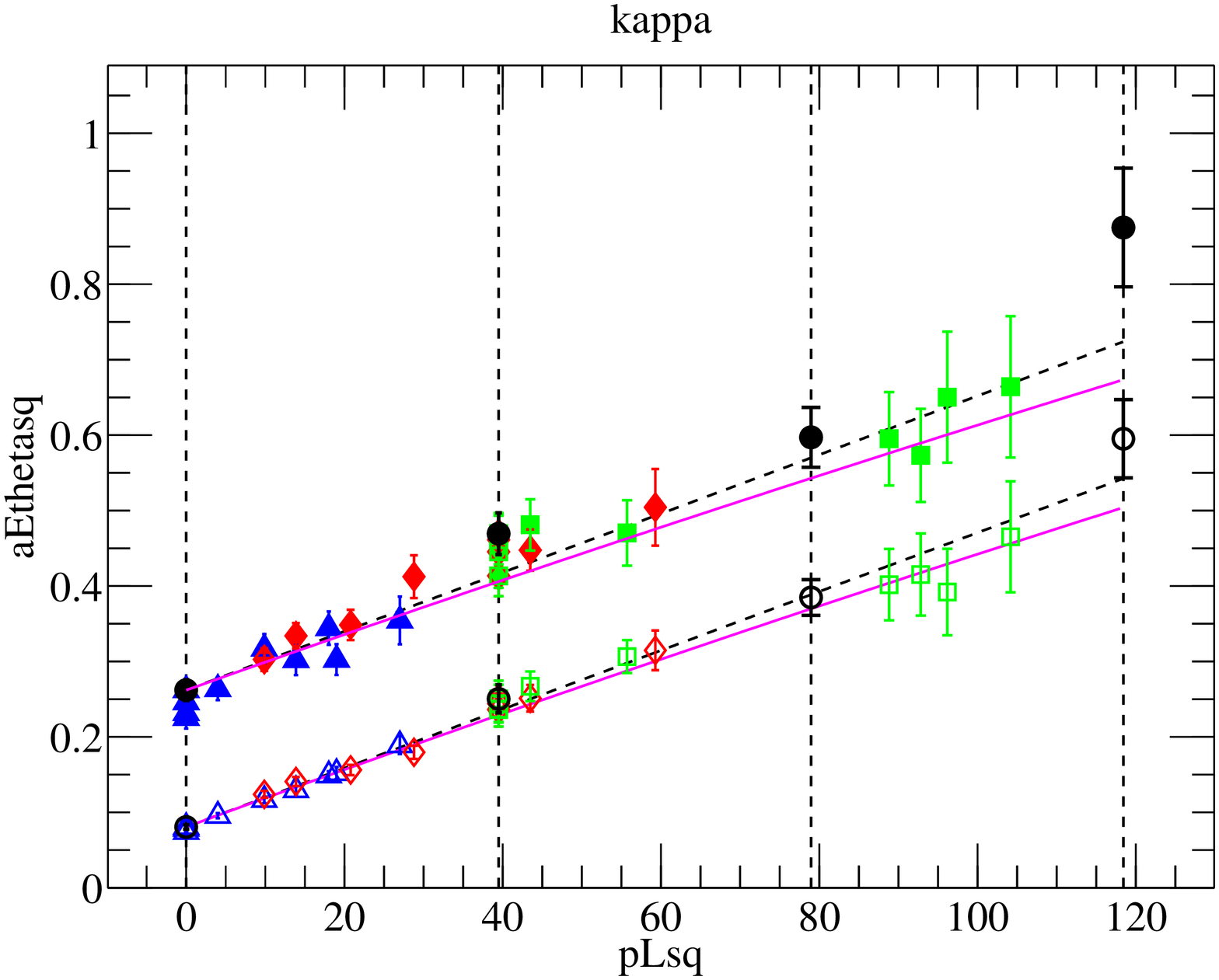}
 \end{minipage}\\[4ex]
 \begin{minipage}{.495\linewidth}
  \psfrag{kappa}[c][t][1][0]{$m_\pi/m_\rho=0.70$}
   \psfrag{relativeerror}[b][b][1][0]{$\delta_{E_{\pi}}$}
   \psfrag{pLsq}[t][c][1][0]{$(\vec{p}L)^2$}
  \epsfig{scale=.275,angle=-90,file=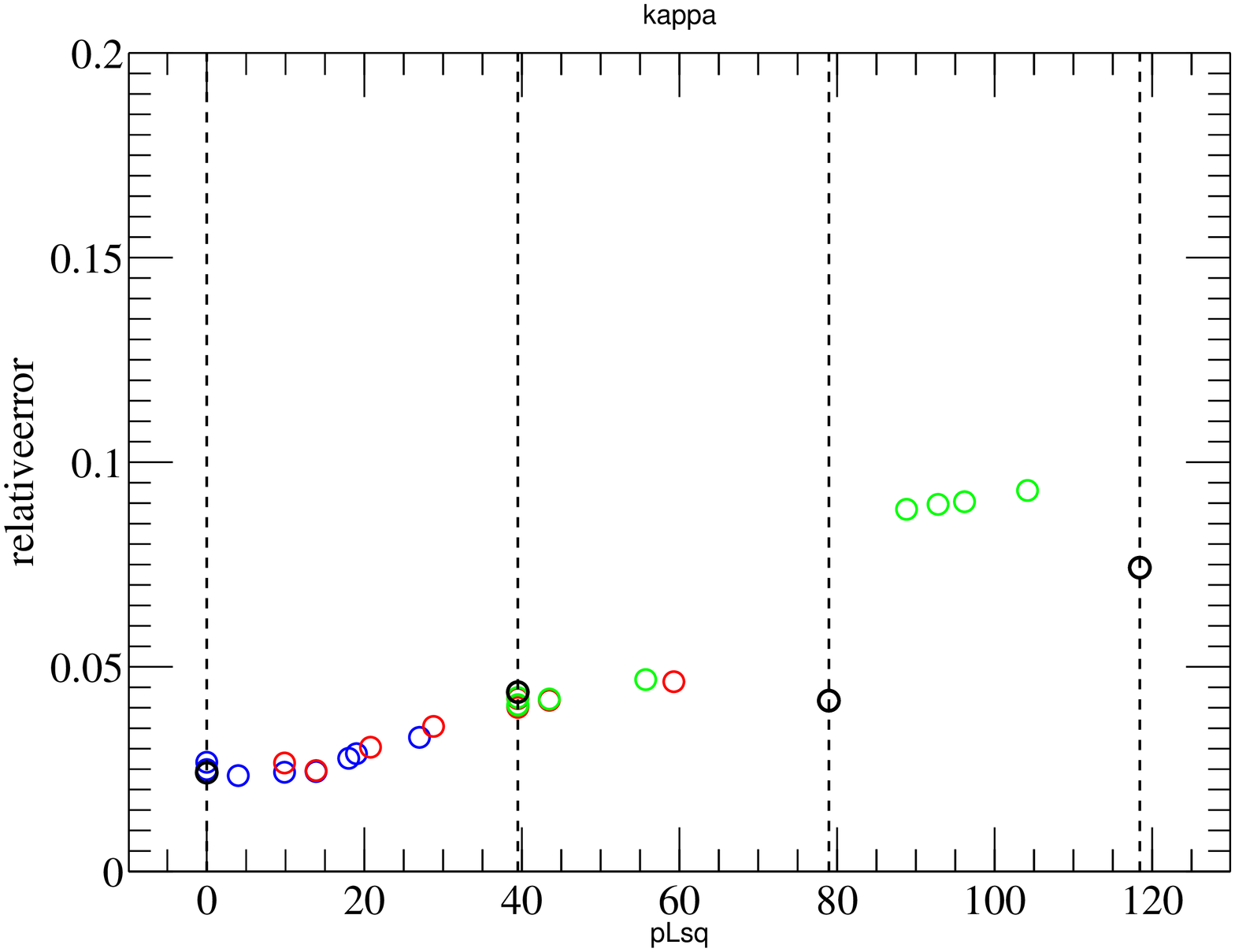}
 \end{minipage}
 \begin{minipage}{.495\linewidth}
   \psfrag{kappa}[c][t][1][0]{$m_\pi/m_\rho=0.57$}
   \psfrag{relativeerror}[b][b][1][0]{$\delta_{E_{\pi}}$}
   \psfrag{pLsq}[t][c][1][0]{$(\vec{p}L)^2$}
  \epsfig{scale=.275,angle=-90,file=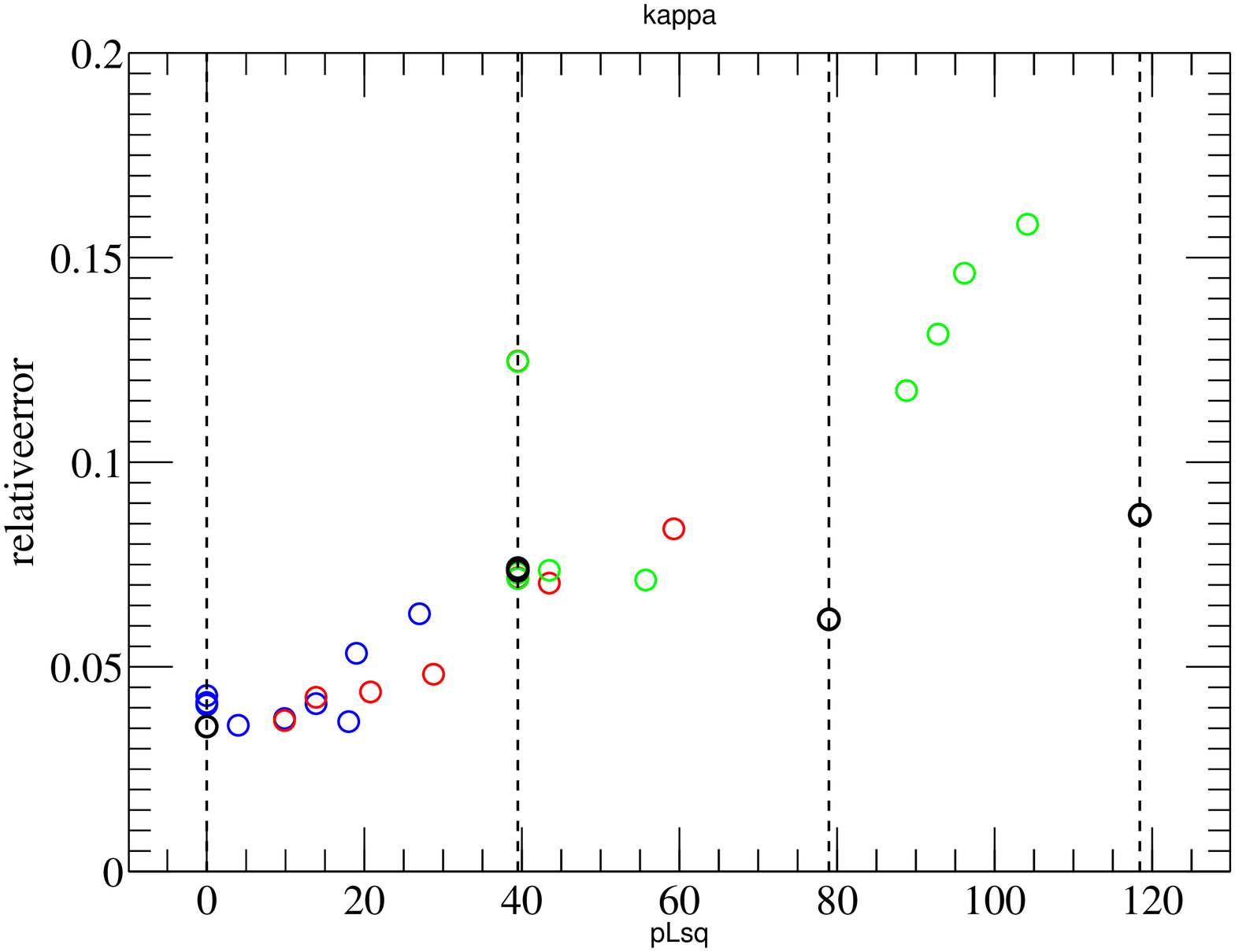}
 \end{minipage}\\[4ex]
\end{center}
\caption{The plots in the first line illustrate the results for the
  dispersion relation for the $\pi$ and the $\rho$ (empty and full
  symbols respectively) for the two choices of the quark mass. In the
  second line we show the corresponding relative error as a function
  of the momentum.} \label{figE}
\end{figure}

\begin{figure}
\begin{center}
 \begin{minipage}{.495\linewidth}
   \psfrag{kappa}[c][t][1][0]{$m_\pi/m_\rho=0.70$}
   \psfrag{pLsq}[t][c][1][0]{$(\vec{p}L)^2$}
   \psfrag{aEthetasq}[b][b][1][0]{$(aE_{\pi/\rho})^2$}
  \epsfig{scale=.275,angle=-90,file=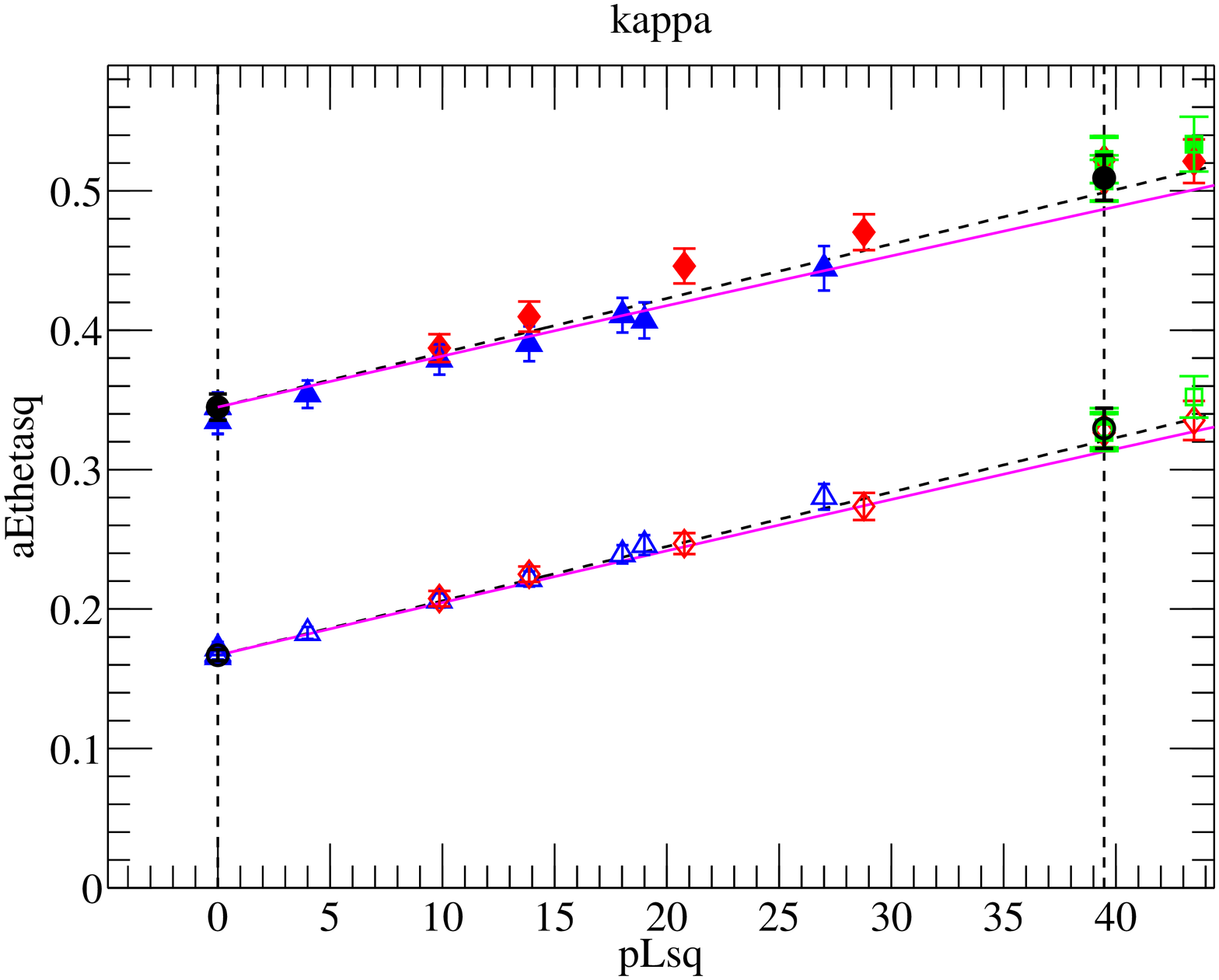}
 \end{minipage}
 \begin{minipage}{.495\linewidth}
   \psfrag{kappa}[c][t][1][0]{$m_\pi/m_\rho=0.57$}
   \psfrag{pLsq}[t][c][1][0]{$(\vec{p}L)^2$}
   \psfrag{aEthetasq}[b][b][1][0]{$(aE_{\pi/\rho})^2$}
  \epsfig{scale=.275,angle=-90,file=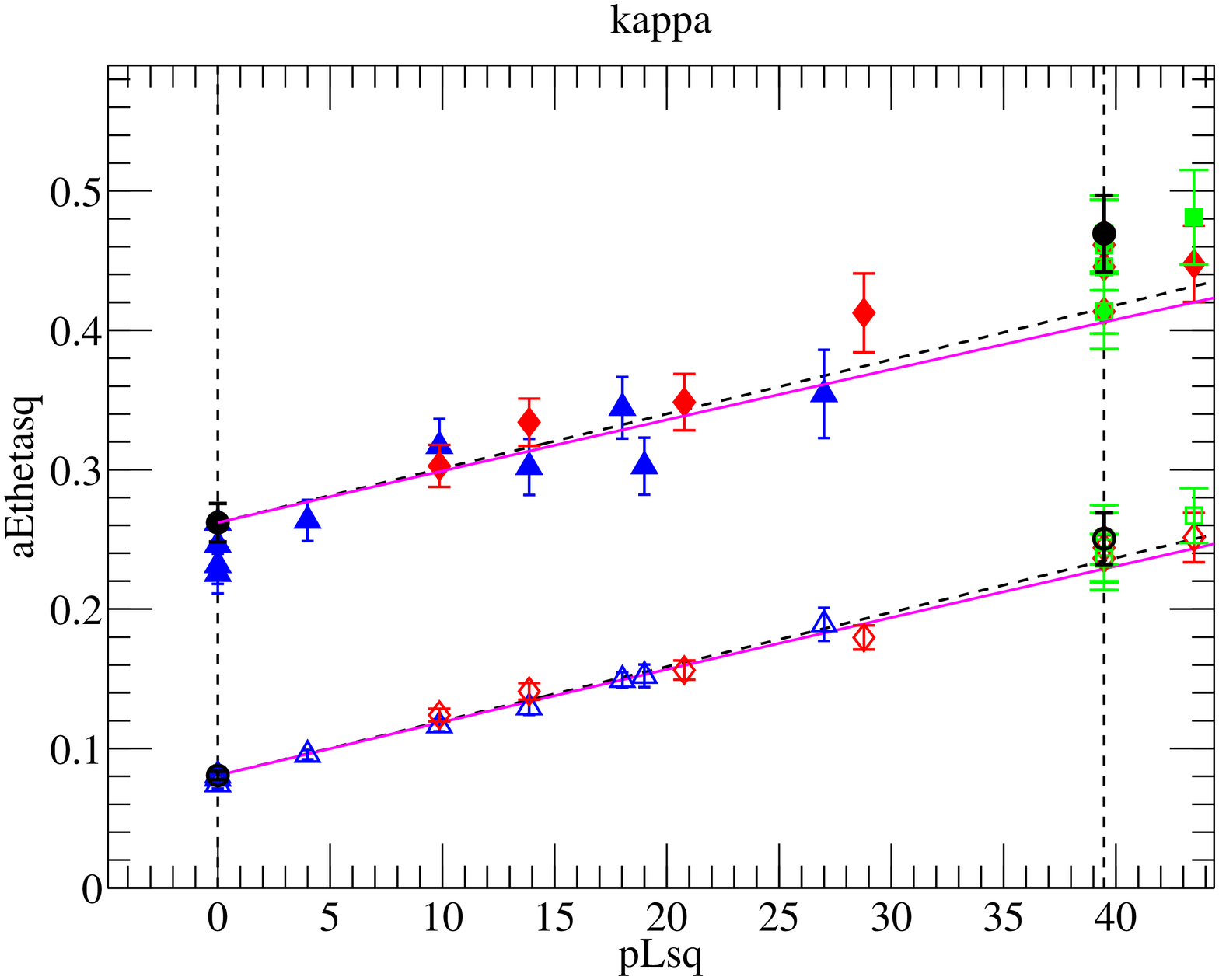}
 \end{minipage}\\[4ex]
\end{center}
\caption{Magnified view of the dispersion relation of fig.~\ref{figE}
  in the interval $|\vec{p}_{\rm lat}|\in [0,2\pi]$.}
\label{figEzoom}
\end{figure}

\begin{figure}
\begin{center}
 \begin{minipage}{.495\linewidth}
   \psfrag{kappa}[c][t][1][0]{$m_\pi/m_\rho=0.70$}
   \psfrag{pLsq}[t][c][1][0]{$(\vec{p}L)^2$}
   \psfrag{Fbare}[b][b][1][0]{$af_{\pi/\rho}$}
  \epsfig{scale=.285,angle=-90,file=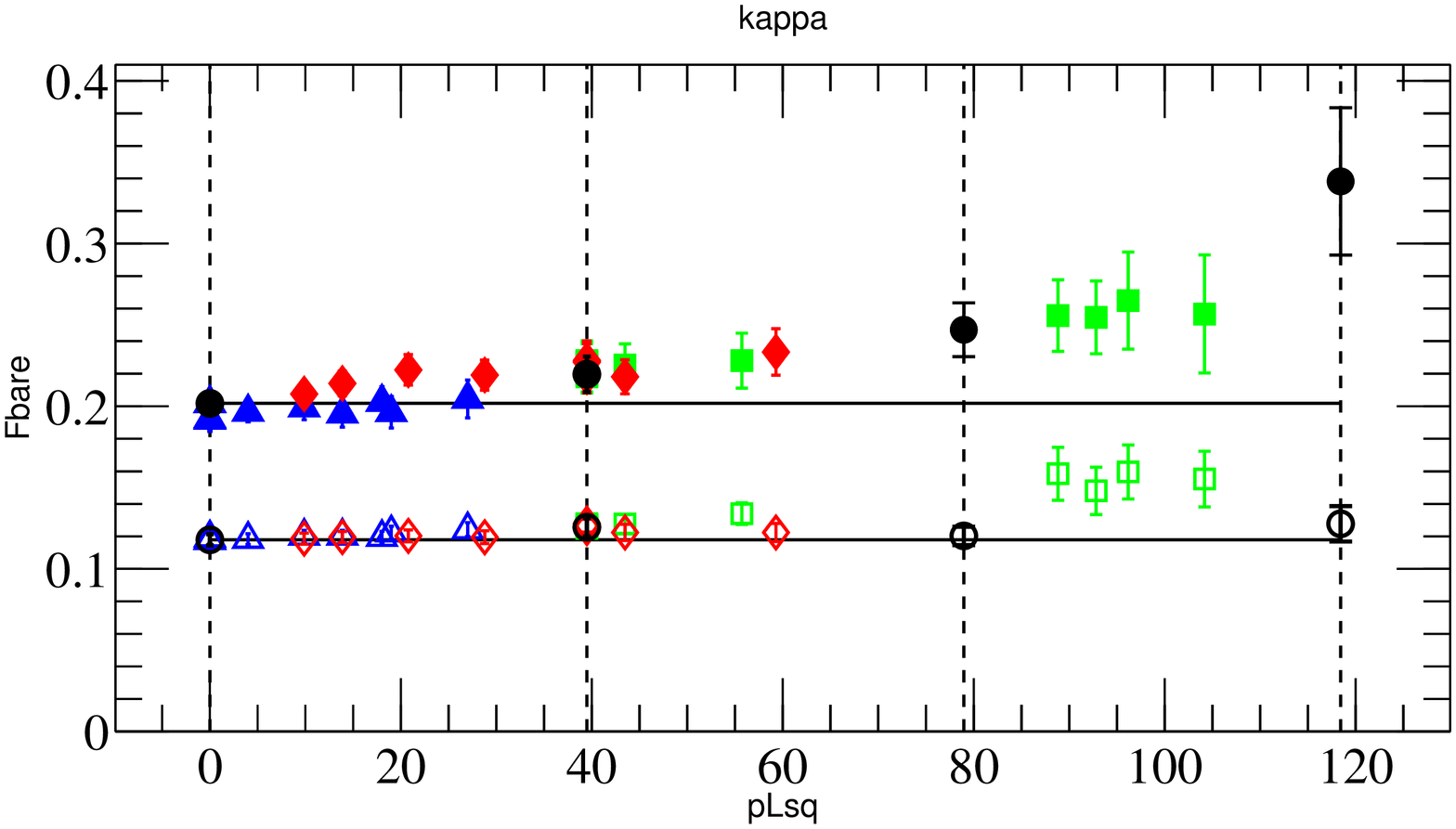}
 \end{minipage}
 \begin{minipage}{.495\linewidth}
   \psfrag{kappa}[c][t][1][0]{$m_\pi/m_\rho=0.57$}
   \psfrag{pLsq}[t][c][1][0]{$(\vec{p}L)^2$}
   \psfrag{Fbare}[b][b][1][0]{$af$}
   \psfrag{Fbare}[b][b][1][0]{$af_{\pi/\rho}$}
   \epsfig{scale=.285,angle=-90,file=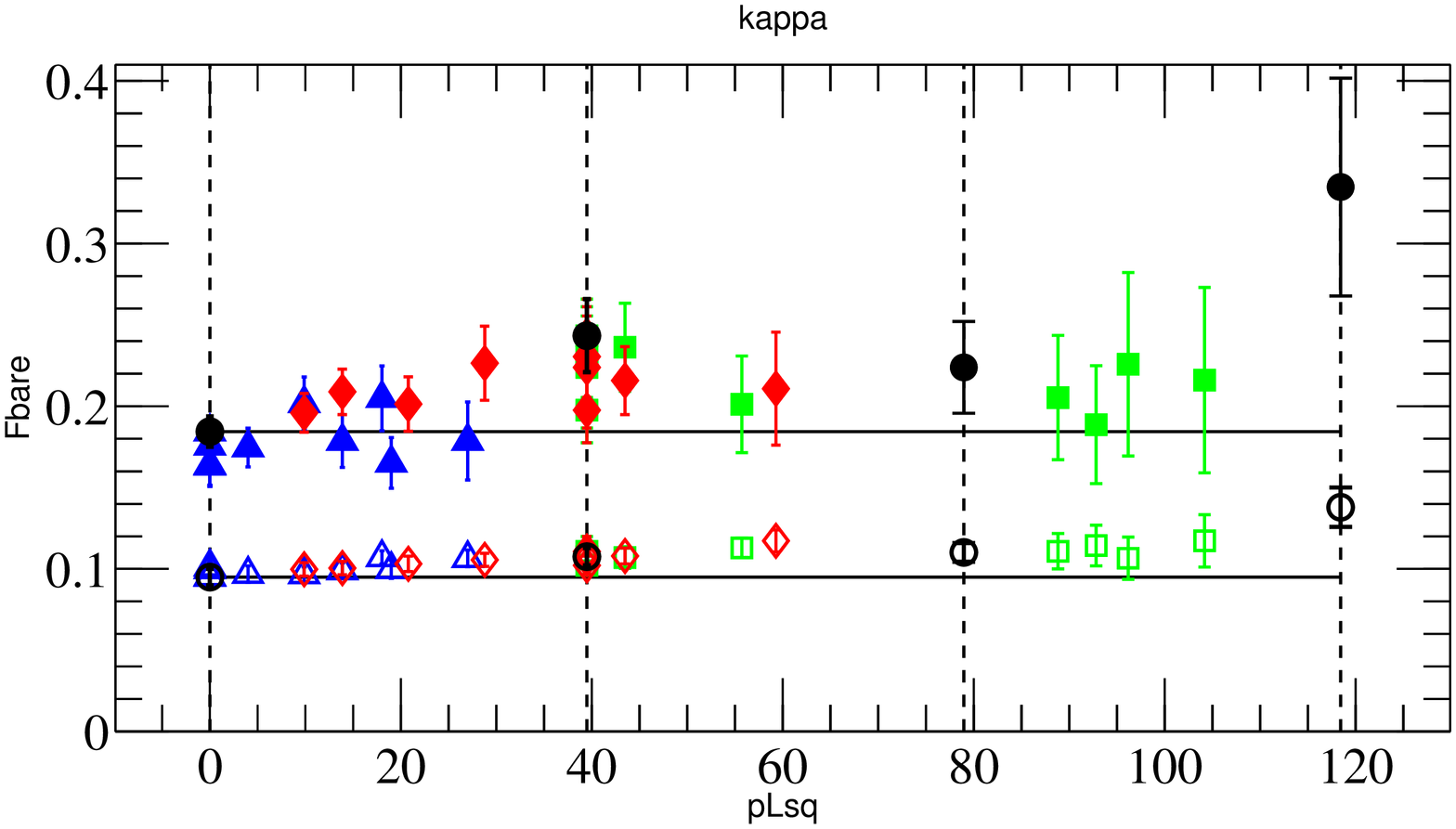}
 \end{minipage}\\[4ex]
 \begin{minipage}{.495\linewidth}
   \psfrag{kappa}[c][t][1][0]{$m_\pi/m_\rho=0.70$}
   \psfrag{pLsq}[t][c][1][0]{$(\vec{p}L)^2$}
   \psfrag{ZP}[c][b][1][0]{$Z_P$}
  \epsfig{scale=.285,angle=-90,file=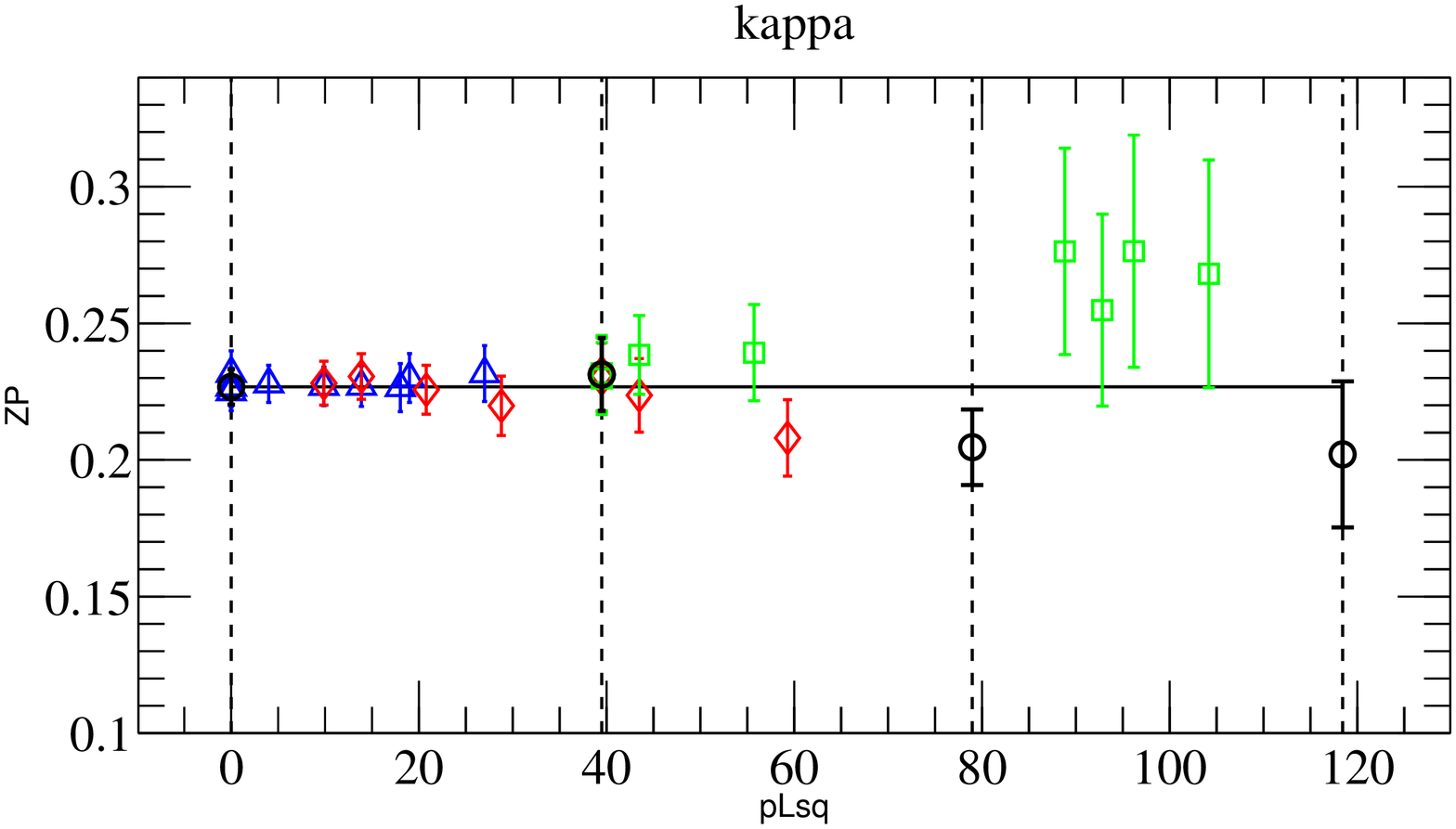}
 \end{minipage}
 \begin{minipage}{.495\linewidth}
   \psfrag{kappa}[c][t][1][0]{$m_\pi/m_\rho=0.57$}
   \psfrag{pLsq}[t][c][1][0]{$(\vec{p}L)^2$}
   \psfrag{ZP}[c][b][1][0]{$Z_P$}
  \epsfig{scale=.285,angle=-90,file=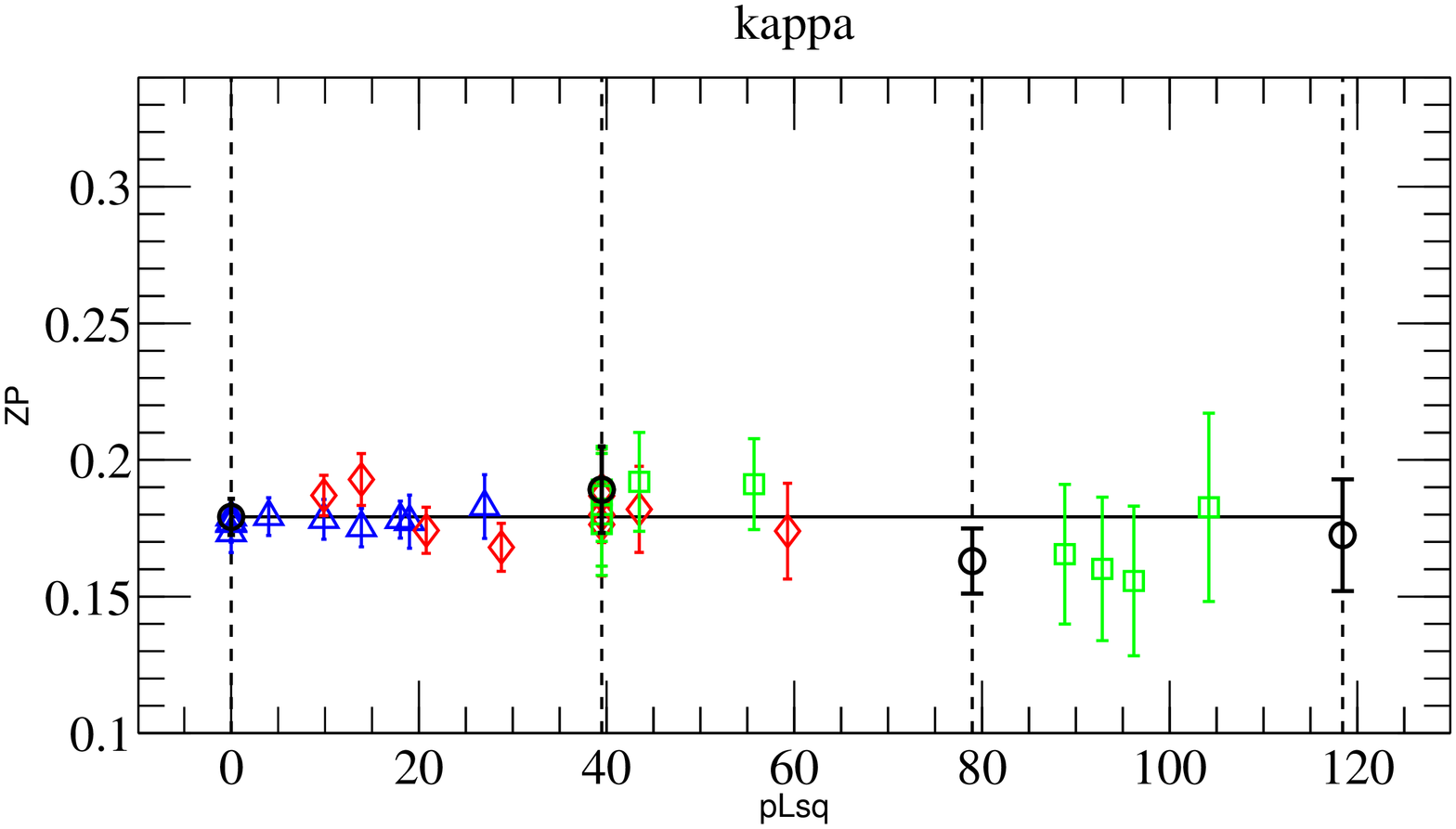}
 \end{minipage}\\[4ex]
\end{center}
\caption{The first line shows the results for the $\pi$ and $\rho$
  decay constant (empty and full symbols respectively) and the second
  line shows the matrix element~(\ref{Z_P}) for the two choices of the
  quark mass. In each plot the horizontal lines represent the central
  value at $\vec p_\mathrm{lat} = \vec\theta_1 = \vec\theta_2 = 0$.}
\label{fig_ZP}
\end{figure}

The series of plots in figures~\ref{figE} and~\ref{fig_ZP} show
our data as a function of $(\vec p L)^2$ in the range $|\vec p
L|\in[0,\sqrt{3}\times2\pi]$. Fig.~\ref{figE} contains the results
for the energies as a function of momentum and fig.~\ref{fig_ZP}
those for the decay constants and $Z_P$. To ease orientation, the
positions of the discrete Fourier momenta $|\vec p_{\rm lat}
L|=0$, $2\pi$, $\sqrt{2}\times2\pi$ and $\sqrt{3}\times2\pi$ are
indicated by dashed vertical lines. We emphasize that it is only
at these values of momenta that one can obtain results using
periodic boundary conditions. In fig.~\ref{figEzoom} we zoom into
the region $|\vec{p}\,L|\le2\pi$ for the dispersion relations. In
this region we would expect lattice artefacts to be small and the
use of twisted boundary conditions to be particularly useful.

In each plot, the (blue) triangles correspond to points in which the
correlation function was evaluated with $\vec{p}_\mathrm{lat}=\vec 0$,
but with all possible pairs of $\vec\theta_1$ and $\vec\theta_2$ from
the set in (\ref{eq:thetas}). The (red) diamonds and (green) squares
represent the results obtained with
$\vec{p}_\mathrm{lat}=(0,2\pi/L,0)$ and
$\vec{p}_\mathrm{lat}=-(0,2\pi/L,0)$ respectively, combined with all
possible pairs of $\vec\theta_1$ and $\vec\theta_2$. The four points
with $\vec\theta_1=\vec\theta_2=\vec 0$ with
$|\vec{p}_\mathrm{lat}|=0,\ 2\pi/L,\ \sqrt{2}\times2\pi/L$ and
$\sqrt{3}\times2\pi/L$ are denoted by (black) circles.

%%%%%%%%%%%%%
\begin{table}
\begin{small}
\begin{center}
\begin{tabular}{l@{\hspace{0.5em}}cccccc}
\hline
\hline\\[-1.5ex]
  &\multicolumn{2}{c}{$\kappa = 0.13500$}
  &\multicolumn{2}{c}{$\kappa = 0.13550$}\\[0.5ex]
  &$\pi$ &$\rho$ &$\pi$ &$\rho$ \\[0.5ex]
\hline\\[-1.5ex]
$\chi^2/\mathrm{d.o.f}|_\mathrm{(\ref{expecteddisprel2})}$
& 0.3& 1.0& 0.6& 1.7\\
$\chi^2/\mathrm{d.o.f}|_\mathrm{(\ref{expecteddisprel1})}$
& 1.8& 2.5& 0.9& 2.1\\[0.5ex]
$\Delta^2$ from~(\ref{expecteddisprel2}) &0.0040(1) &0.0042(2)
&0.0040(1)
&0.0048(4) \\[0.5ex]
\hline \hline
\end{tabular}
\caption{$\chi^2$/d.o.f. for the lattice data with respect the
  expectations eqs.~(\ref{expecteddisprel2})
  and~(\ref{expecteddisprel1}) with $\Delta^2=a^2/L^2=0.0039$ (first
  two rows) and the results obtained from a fit
  to~(\ref{expecteddisprel2}) with $\Delta^2$ left as a parameter of
  the fit (third row).}
\label{tab_slopes}
\end{center}
\end{small}
\end{table}

%%%%%%%%%%%%%%%%%%

For the discussion of our results it is convenient to rewrite the
dispersion relation in eq.\,(\ref{eq:disprel}) in the form
\begin{equation}
(aE_{\pi/\rho})^2= (am_{\pi/\rho})^2 + \Delta^2(\vec{p}\, L)^2
\label{expecteddisprel2}
\end{equation}
where $\Delta^2=(a/L)^2 = 0.0039$. The dispersion
relation~(\ref{expecteddisprel2}) is displayed as the dashed line in
the plots of fig.~\ref{figE}. In the first row of
table~\ref{tab_slopes} we present the $\chi^2/$d.o.f of the comparison
of our data to eq.~(\ref{expecteddisprel2}) over the range
$0\le|\vec{p}\,|^2L^2\le(2\pi)^2$ using the values of the meson masses
obtained from fits at zero momenta. In the third row of the table we
present the values of $\Delta^2$ obtained by fitting the lattice data
to the functional form in eq.~(\ref{expecteddisprel2}) over the same
range in momentum, but allowing $\Delta^2$ to be a parameter of the
fit. 
We note that our values for the ratios $m_\pi/m_\rho$ agree with those
found earlier on the same configurations in~\cite{Allton:2001sk}.

As the momentum of the meson grows so do the expected discretization
effects in the dispersion relation. For example, a free scalar
particle with a wave-function $\phi(x)$ satisfying the
(Minkowski-Space) Klein-Gordon equation $(\square+m^2)\phi(x)=0$, with
a generic discretized second derivative defined by
$\partial^2f(x)/\partial x^2 = (f(x+a)+f(x-a)-2f(x))/a^2$, satisfies
the following lattice dispersion relation:
\begin{equation}
\sinh^2\bigg({aE_{\pi/\rho}\over2}\bigg) =
    \sinh^2\bigg({am\over 2}\bigg)
  + \sum_i\sin^2\bigg(\Delta{p_{i}L\over 2}\bigg).
  \label{expecteddisprel1}
\end{equation}
To indicate the possible size of lattice artefacts we plot the
dispersion relation of eq.~(\ref{expecteddisprel1}) as the solid curve
in fig.~\ref{figE}. We stress, however, that in an interacting theory
the discretization errors will in general be different from those in
eq.~(\ref{expecteddisprel1}). Indeed there seems to be no evidence
from our data that eq.~(\ref{expecteddisprel1}) is a particularly good
representation of the lattice artefacts (see the second row of
table~\ref{tab_slopes}, where we show the $\chi^2/$d.o.f. from a
comparison of our data with~(\ref{expecteddisprel1})). At small
momenta the solid curve merges of course with the dashed line
representing the continuum dispersion relation.

We conclude from the results for the dispersion relation plotted in
figs.~\ref{figE} and~\ref{figEzoom} that the use of partially twisted
boundary conditions is beautifully consistent with expectations,
particularly at low momenta where the lattice artefacts are
small\footnote{The different data points at $|\vec p L|=0$ and
  $2\pi$ correspond to $\vec\theta_1=\vec\theta_2$ but with different
  choices of $\vec\theta_1$ and $\vec\theta_2$.}. An important further
observation is that there is no evidence in our data that the
introduction of twisted boundary conditions increases significantly
the statistical or systematic uncertainties. This is illustrated in
the second line of figure~\ref{figE}, which shows the relative error
in the pion energy
\begin{equation}
\delta_{E_{\pi}}\equiv{\delta E_{\pi}\over E_{\pi}}
\end{equation}
as $|\vec{p}\, L|$ is varied. The error $\delta E_{\pi}$ is the
jackknife error, including the statistical error and the systematic
uncertainty stemming from the improvement constants in the improved
quark currents. The plot shows that the errors increase smoothly as
the momentum increases, and that no appreciable additional noise is
introduced by partial twisting
\footnote{For $\kappa=0.13550$ at $|\vec pL|=\tpol$ we observe a fluctuation in the effective mass from one of the gauge configurations and with our choice of the position of the source (this fluctuation has also been observed by our colleagues in the UKQCD collaboration \cite{pricomMcNeile}). The fluctuation is particularly noticeable when twisting both quarks by  $\vec\theta =
(3,3,3)$ and this is the reason for the larger jackknife error at this particular momentum and twist combination (see figure~\ref{figE}).}.
We observe the same behaviour for all analyzed quantities.

In fig.~\ref{fig_ZP} we plot our results for the decay constants
$f_\pi$ and $f_\rho$ and for $Z_P$.
The values for $af_\pi$ agree
with the ones obtained in \cite{Allton:2004qq} at 
$|\vec{p}\, L|=0$. Again we see that the 
results are completely consistent with theoretical expectations, being
independent of the twisting angles and Fourier momenta.

%%%%%%%%%%%%%%%%%%%%%%%%%%%%%%%%%%%%%%
\section{Conclusions}\label{sec:concs}
%%%%%%%%%%%%%%%%%%%%%%%%%%%%%%%%%%%%%%
\vspace{-.2cm}

We have investigated the use of partially twisted boundary conditions
in evaluating the energies of pseudo scalar and vector mesons and their
leptonic decay constants. The results are very encouraging; it does
appear that the method allows the evaluation of physical quantities
with any momentum. Moreover the use of these boundary conditions does
not appear to increase the errors in any appreciable way. It will be
important to monitor whether this continues to be true as the quark
masses are decreased. Once the quark masses are such that two-pion
intermediate states contribute significantly to the $\rho$-meson's
correlation function the finite-volume effects will no longer fall
exponentially with the volume, but only as powers. For the pion
observables studied in this paper this is not the case.

Partially twisted boundary conditions will be particularly useful for
evaluating momentum-dependent physical quantities. One important
application is to the determination of the form-factors of
semileptonic weak decays of heavy ($D$ and $B$) mesons to light
mesons. For these processes, with conventional periodic boundary
conditions, the initial and final state hadrons are restricted to have
momenta $(\tpol)\vec{n}$ where $\vec{n}$ is a vector of integers. In
order to avoid lattice artefacts the possible values of $|\vec{n}|$
are frequently limited to $0$, $1$, and perhaps $\sqrt{2}$. Thus, for
any particular choice of quark masses, the number of values of the
momentum transfer, $q^2$, or the light meson energy, $E$, is also very
limited. Moreover, chiral extrapolations are conveniently performed at
fixed $q^2$~\cite{Bowler:1999xn,Abada:2000ty,Bowler:2004zb} or fixed
$E$~\cite{Aoki:2001rd,Aubin:2004ej,Shigemitsu:2004ft,Okamoto:2004xg}
(and heavy quark extrapolations at fixed $E$), but $q^2$ and $E$ vary
with both the momentum and quark masses. Ans\"atze for the form
factors, such as the Becirevic-Kaidalov~\cite{Becirevic:1999kt} model,
are used to interpolate and extrapolate simulation data to sets of
common $q^2$ or $E$ values before the extrapolations are performed.
Using twisted boundary conditions would enable the form factors to be
evaluated directly at these common values, removing the need for the
intermediate form-factor fit.

For some other physical quantities, such as the moments of hadronic
deep inelastic structure functions or light-cone distribution
amplitudes, it may be helpful to use twisted boundary conditions even
though it is not strictly necessary. The corresponding matrix elements
are proportional to factors of $p_i$, where $\vec{p}$ is the momentum
of the hadron, so that the correlation functions must be computed with
$\vec{p}\neq 0$. The use of twisted boundary conditions allows
$|\vec{p}\,|$ to be decreased and hence the lattice artefacts to be
reduced. Moreover by varying $\vec{p}$ one can verify that the leading
twist component has been extracted correctly.

Following the successful conclusion of this exploratory numerical
study of the implementation of partially twisted boundary conditions
we now look forward to applying them in lattice computations of a wide
variety of phenomenologically important quantities.

{\bf Acknowledgements}
We warmly thank Daragh Byrne, Steve Downing and Craig McNeile for
their support with QCDgrid and Massimo di Pierro for help using FermiQCD. 
We thank the Iridis parallel
computing team at the University of Southampton, in particular Oz
Parchment and Ivan Wolton, for their assistance. We also acknowledge
Alan Irving, Craig McNeile and Chris Michael for correspondence on the
gauge field configurations. 
This work was
supported by PPARC grants PPA/G/S/2002/00467 and
PPA/G/O/2002/00468.

\bibliographystyle{elsevier}
\vspace{-.5cm}
\bibliography{twisted_numerical}

\begin{thebibliography}{10}

\bibitem{Bedaque:2004kc}
P.F. Bedaque, Phys. Lett. B593 (2004) 82, nucl-th/0402051.
%%CITATION = NUCL-TH 0402051;%%

\bibitem{deDivitiis:2004kq}
G.M. de~Divitiis, R.~Petronzio and N.~Tantalo, Phys. Lett. B595 (2004) 408,
  hep-lat/0405002.
%%CITATION = HEP-LAT 0405002;%%

\bibitem{Sachrajda:2004mi}
C.T. Sachrajda and G.~Villadoro, Phys. Lett. B609 (2005) 73, hep-lat/0411033.
%%CITATION = HEP-LAT 0411033;%%

\bibitem{Bedaque:2004ax}
P.F. Bedaque and J.W. Chen  (2004), hep-lat/0412023.
%%CITATION = HEP-LAT 0412023;%%

\bibitem{Allton:2001sk}
UKQCD Collaboration, C.R. Allton et~al., Phys. Rev. D65 (2002) 054502,
  hep-lat/0107021.
%%CITATION = HEP-LAT 0107021;%%

\bibitem{Allton:2004qq}
UKQCD Collaboration, C.R. Allton et~al., Phys. Rev. D70 (2004) 014501,
  hep-lat/0403007.
%%CITATION = HEP-LAT 0403007;%%

\bibitem{DiPierro:2000bd}
M.~Di~Pierro, Comput. Phys. Commun. 141 (2001) 98, hep-lat/0004007.
%%CITATION = HEP-LAT 0004007;%%

\bibitem{DiPierro:2001yu}
M.~Di~Pierro, Nucl. Phys. Proc. Suppl. 106 (2002) 1034, hep-lat/0110116.
%%CITATION = HEP-LAT 0110116;%%

\bibitem{DiPierro:2003sz}
FermiQCD Colaboration Collaboration, M.~Di~Pierro et~al., Nucl. Phys. Proc.
  Suppl. 129 (2004) 832, hep-lat/0311027.
%%CITATION = HEP-LAT 0311027;%%

\bibitem{Harada:2002jh}
J.~Harada, S.~Hashimoto, A.S. Kronfeld and T.~Onogi, Phys. Rev. D67 (2003)
  014503, hep-lat/0208004.
%%CITATION = HEP-LAT 0208004;%%

\bibitem{DellaMorte:2005se}
M.~Della~Morte, R.~Hoffmann and R.~Sommer, JHEP 03 (2005) 029, hep-lat/0503003.
%%CITATION = HEP-LAT 0503003;%%

\bibitem{pricomMcNeile}
A.~Irving, C.~McNeile and C.~Michael, private communications.

\bibitem{Bowler:1999xn}
UKQCD Collaboration, K.C. Bowler et~al., Phys. Lett. B486 (2000) 111,
  hep-lat/9911011.
%%CITATION = HEP-LAT 9911011;%%

\bibitem{Abada:2000ty}
A.~Abada et~al., Nucl. Phys. B619 (2001) 565, hep-lat/0011065.
%%CITATION = HEP-LAT 0011065;%%

\bibitem{Bowler:2004zb}
UKQCD Collaboration, K.C. Bowler, J.F. Gill, C.M. Maynard and J.M. Flynn, JHEP
  05 (2004) 035, hep-lat/0402023.
%%CITATION = HEP-LAT 0402023;%%

\bibitem{Aoki:2001rd}
JLQCD Collaboration, S.~Aoki et~al., Phys. Rev. D64 (2001) 114505,
  hep-lat/0106024.
%%CITATION = HEP-LAT 0106024;%%

\bibitem{Aubin:2004ej}
Fermilab Lattice Collaboration, C.~Aubin et~al., Phys. Rev. Lett. 94 (2005)
  011601, hep-ph/0408306.
%%CITATION = HEP-PH 0408306;%%

\bibitem{Shigemitsu:2004ft}
J.~Shigemitsu et~al., Nucl. Phys. Proc. Suppl. 140 (2005) 464, hep-lat/0408019.
%%CITATION = HEP-LAT 0408019;%%

\bibitem{Okamoto:2004xg}
M.~Okamoto et~al., Nucl. Phys. Proc. Suppl. 140 (2005) 461, hep-lat/0409116.
%%CITATION = HEP-LAT 0409116;%%

\bibitem{Becirevic:1999kt}
D.~Becirevic and A.B. Kaidalov, Phys. Lett. B478 (2000) 417, hep-ph/9904490.
%%CITATION = HEP-PH 9904490;%%

\end{thebibliography}

\end{document}